\documentclass[10pt,twocolumn,aps,pra,longbibliography]{revtex4-2}

\usepackage{physics}
\usepackage{graphicx}
\usepackage[dvipsnames,svgnames,table]{xcolor}
\usepackage[unicode=true,
            pdfusetitle,
            bookmarks=true,
            bookmarksnumbered=false,
            bookmarksopen=false,
            breaklinks=true,
            pdfborder={0 0 0},
            backref=false,
            colorlinks=true,
            hypertexnames=false]{hyperref}
\hypersetup{linkcolor=NavyBlue,urlcolor=NavyBlue,citecolor=NavyBlue}

\begin{document}
\title{All-optical correlated noisy channel and its application in recovering quantum coherence}

\author{Dan Lei}
\affiliation{Department of Physics, Hangzhou Dianzi University, Hangzhou 310018, China}
\author{Disheng Guo}
\affiliation{Department of Physics, Hangzhou Dianzi University, Hangzhou 310018, China}
\author{Jun Xin}
\email{jxin@hdu.edu.cn}
\affiliation{Department of Physics, Hangzhou Dianzi University, Hangzhou 310018, China}
\author{Xiao-Ming Lu}
\email{lxm@hdu.edu.cn}
\affiliation{Department of Physics, Hangzhou Dianzi University, Hangzhou 310018, China}

\begin{abstract}
	Attenuation and amplification are the most common processes for optical communications.
	Amplification can be used to compensate the attenuation of the complex amplitude of an optical field, but is unable to recover the coherence lost, provided that the attenuation channel and the amplification channel are independent.
	In this work, we show that the quantum coherence of an optical filed can be regained if the attenuation channel and the amplification channel share correlated noise.
	We propose an all-optical correlated noisy channel relying on four-wave mixing process and demonstrate its capability of recovering quantum coherence within continuous-variable systems. 
	We quantitatively investigate the coherence recovery phenomena for coherent states and two-mode squeezed states. 
	Moreover, we analyze the effect of other photon losses that are independent with the recovery channel on the performance of recovering coherence. 
	Different from correlated noisy channels previously proposed based on electro-optic conversions, the correlated noisy channel in our protocol is all-optical and thus owns larger operational bandwidths.
\end{abstract}

\maketitle

\section{Introduction}
Quantum coherence is a basic feature that marks the departure of quantum realm from the classical world~\cite{PhysRevLett.113.140401,RevModPhys.89.041003,Hu2018}.
Arising from the superposition principle, quantum coherence embodies the essence of quantum correlations including quantum entanglement~\cite{PhysRevLett.115.020403,PhysRevLett.116.240405,PhysRevLett.117.020402} and quantum steering~\cite{PhysRevA.95.010301}. 
As a resource for information processing~\cite{PhysRevLett.113.140401,RevModPhys.89.041003,Hu2018}, quantum coherence is fragile as decoherence inevitably occurs due to the interaction between a quantum system and its environment. 
Much effort has been devoted to mitigating the decoherence in discrete variable quantum systems, with notable examples including dynamical decoupling~\cite{PhysRevLett.82.2417,Du2009,Lange2010}, error correcting codes~\cite{PhysRevA.52.R2493,PhysRevLett.77.2585,Bennett1996,Steane1996,Gottesman1996,Knill1997}, reservoir engineering~\cite{PhysRevLett.86.4988}, inversion of quantum jumps~\cite{PhysRevLett.76.3108}, and feedback control~\cite{PhysRevA.75.032323,PhysRevLett.113.020407}.
In addition to the discrete variable regime, continuous variable systems, such as quantum oscillators and optical fields, are also of great significance in quantum information processing~\cite{Braunstein2005}.

Quantum correlation can be used to against the noise effect during quantum information processing.
When the noise channel has memory effect or share correlation~\cite{Lupo2010,Caruso2014}, it is possible to recover some quantum resource that are damaged due to decoherence.
For example, revival of squeezing~\cite{Deng2016a} and Einstein-Podolsky-Rosen (EPR) steering~\cite{Deng2021} have  been observed in the experiment using correlated noisy channels.
Such correlated noisy channels can also be used to implement some information processing tasks that are impossible by  only using independent channels, e.g., Gaussian error correction can be fulfilled via correlated noisy channels~\cite{PhysRevLett.111.180502}.

Up to now, correlated noisy channels are all established based on the conventional feed-forward techniques.
It employs the same signal generators to produce the correlated noise between the quantum system and environment, and then uses electro-optic modulators to perform the encoding and decoding procedures in classical channels~\cite{PhysRevLett.111.180502,Deng2016a,Deng2021}.
However, such conventional correlated noisy channels are limited by the electrical bandwidth of the modulators.
Note that the operational bandwidth is an important factor for correlated noisy channel.
This is because the quantum properties (e.g., quantum coherence) of a quantum system may be distributed within a large range of frequency spectrum.
To recover the quantum properties of a quantum system that passes through a noisy channel, the operational bandwidth of the correlated noisy channel should match with that of the noisy channel.
Therefore, it is valuable to further enhance the operational bandwidth of the correlated noisy channel.
This leads us to explore an all-optical version of the correlated noisy channel.

In this work, we propose an all-optical correlated noisy channel (ACNC) based on four-wave mixing (FWM) processes in hot atomic ensembles and to investigate its capability of recovering quantum coherence.
Different from conventional correlated noisy channels, the ACNC avoids the electro-optic conversions and its noisy channels are all-optical. 
Due to the unique advantage of all-optical strategy~\cite{Hillerkuss2011,Takeda2019}, the ACNC owns larger operational bandwidth than the conventional correlated noisy channels. 
We use the ACNC to recover the coherence loss of an optical field, that is caused by the attenuation modeled by beam splitting.
A common amplification process can compensate the attenuation of the complex amplitude of an optical field but cannot recover the loss of coherence. 
The amplification even brings in excessive noise~\cite{Haus1962,Caves1982,Yamamoto1986,Clerk2010}.
We shall show that utilizing the correlated noise in the attenuation channel and the amplification channel, the ACNC can recovery a portion of coherence while compensating the loss of the complex amplitude.
We use the Gaussian relative entropy of coherence~\cite{PhysRevA.93.032111} as a coherence measure to quantitatively study the coherence recovery for coherent states and two-mode squeezed states (TMSS). 
In addition, we also investigate the effect of imperfect factors on the performance of the ACNC, such as the photon absorption occurring in the atomic cells and unavoidable losses in light path.

This paper is organized as follows.
In Sec.~\ref{sec:ACNC}, we give the design of the ACNC and all the relevant input-output relations for the processes therein.
In Sec.~\ref{sec:recovery_coherence}, we give a brief review on the coherence measure for Gaussian states and then study the capability of the ACNC for recovering the quantum coherence of a single-mode coherent state and a TMSS. 
We also analyze the the impact of photon loss.
We summarize our work in Sec.~\ref{sec:conclusion}.

\section{ACNC based on FWM processes} \label{sec:ACNC}

\begin{figure}[tb]
	\centering
	\includegraphics[width=3.4in]{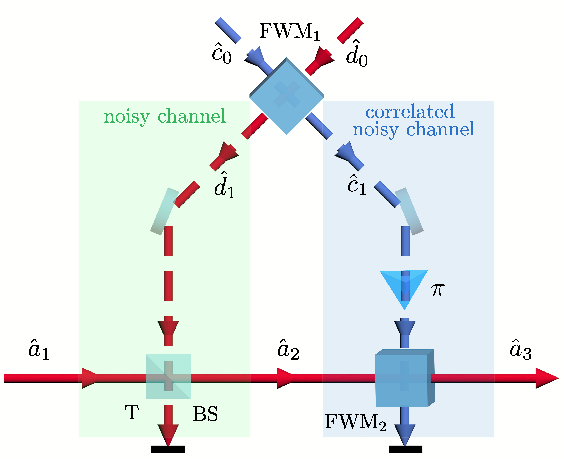}
	\caption{Design of the ACNC based on BS and FWM.}
	\label{fig1}
\end{figure}

Figure~\ref{fig1} illustrates the configuration of ACNC, which employs two FWM processes and a beam splitting process. 
The first one (FWM$_{1}$) generates the correlated noise, while the second one (FWM$_{2}$) serves as a decoder that uses the quantum correlation produced from FWM$_{1}$ to recover the coherence of the quantum system. 
Under the ``undepleted pump'' approximation~\cite{Jasperse}, the interaction Hamiltonian of FWM$_{1}$ is of the form $\hat{H}_{1}=i\hbar\xi_{1}(\hat{c}_{1}\hat{d}_{1} - c_1^\dagger d_1^\dagger) $, where the parameter $\xi_{1}$ is the interaction strength of FWM\(_1\), $\hat{c}_{1}$ and $\hat{d}_{1}$ are the annihilation operators of the corresponding modes, respectively. 
This interaction Hamiltonian guarantees that the photons in modes $\hat{c}_{1}$ and $\hat{d}_{1}$ are produced simultaneously, and therefore strong quantum correlations will be generated between these two modes. 
The input-output relationship of FWM$_{1}$ is
\begin{align}
	\hat{c}_{1} &= G_{1}\hat{c}_{0}+g_{1}\hat{d}^{\dagger}_{0}, \label{eq:c_1}\\
	\hat{d}_{1} &= G_{1}\hat{d}_{0}+g_{1}\hat{c}^{\dagger}_{0}, \label{eq:d_1}
\end{align}
where \(c_0\) and \(d_0\) are the annihilation operators for the two input modes of FWM$_{1}$, respectively, $G_{1}=\cosh{(\xi_{1}\tau)}$ is the amplitude gain of FWM$_{1}$ with $\tau$ being the interaction timescale, and $g_{1} \equiv \sqrt{G^{2}_{1}-1}$.
When the input of FWM$_{1}$ are in vacuum, its output state is known as the EPR entangled state~\cite{Boyer2008,Marino2009}. 
It has been shown that both the position and momentum quadratures of either $\hat{c}_{1}$ or $\hat{d}_{1}$ yield thermal noise. 
However, we can use the EPR correlation to reduce such thermal noise via a joint homodyne measurement~\cite{PhysRevLett.68.3663}. 

We now consider a single-mode Gaussian optical field, denoted as $\hat{a}_{1}$, that passes through a noisy channel that is realized by mixing $\hat{a}_{1}$ with $\hat{d}_{1}$ on a linear beam splitter (BS) with the transmissivity $T$.
The input-output relation of the BS is 
\begin{align}
	\hat{a}_{2} &= \sqrt{T}\hat{a}_{1}+\sqrt{1-T}\hat{d}_{1}, \label{a_2} \\
	\hat{d}_{2} &= \sqrt{1-T}\hat{a}_{1} - \sqrt{T}\hat{d}_{1}, \label{eq:d_2}
\end{align}
where \(\hat a_2\) is taken as the output of the noisy channel and the other output mode \(\hat d_2\) of the BS will not be considered. 
After the BS, the noise owned by $\hat{d}_{1}$ is injected into the optical field $\hat{a}_{1}$ and thus may destroy the coherence of the input quantum state. 

To recovery the Gaussian information of \(a_1\), we then use $\hat{a}_{2}$ as the input mode of FWM$_{2}$, while the other port of FWM$_{2}$ is seeded by the mode $\hat{c}_{1}$ with a phase delay \(\phi\).
The input-output relation of FWM$_{2}$ is 
\begin{align}
	\hat{a}_3 &= G_2 \hat{a}_2 + g_2 e^{i\phi} \hat{c}^{\dagger}_1, \\
	\hat{c}_2 &= G_2 e^{-i\phi} \hat{c}_1 + g_2 \hat{a}^{\dagger}_2,
\end{align}
where $G_{2}$ is the amplitude gain of FWM$_{2}$ and $g_{2} \equiv \sqrt{G^{2}_{2}-1}$. 
We take \(\hat a_3\) as the final output mode.
Substituting Eqs.~(\ref{eq:c_1}--\ref{a_2}) into Eq.~\eqref{acncoutput}, \(a_3\) can be expressed with respect to the initial modes as 
\begin{align} \label{acncoutput}
	\hat{a}_{3} &= G_{2}\sqrt{T}\hat{a}_{1} + \qty(G_{1}G_{2}\sqrt{1-T}+g_{1}g_{2}e^{i\phi})\hat{d}_{0} \nonumber \\
	& \quad + \qty(G_{2}g_{1}\sqrt{1-T}+G_{1}g_{2}e^{i\phi})\hat{c}^{\dagger}_{0}. 
\end{align}
To compensate the amplitude loss of \(a_1\), we henceforth set $G_{2}=1/\sqrt{T}$ so that \(\ev*{\hat a_3} = \ev*{\hat a_1}\) for \(\ev*{\hat c_0} = \ev*{\hat d_0} = 0\).
Moreover, it can be seen from Eq.~(\ref{acncoutput}) that extra noise is introduced by modes $\hat{d}_{0}$ and $\hat{c}_{0}$. 
To simply this extra noise, we take $\phi=\pi$ throughout this work. 
As a result, Eq.~(\ref{acncoutput}) can be rewritten as
\begin{equation}\label{acncoutput2}
\hat{a}_{3}=\hat{a}_{1}+g_{2}(G_{1}-g_{1})(\hat{d}_{0}-\hat{c}^{\dagger}_{0}).
\end{equation}
Note that the last term in Eq.~(\ref{acncoutput2}) vanishes in the limit of $G_{1}\gg1$, meaning that the additive noise introduced by modes $\hat{c}_{1}$ and $\hat{d}_{1}$ can be approximately cancelled when the intensity gain of FWM$_{1}$ is large enough. 

The above-mentioned noise cancellation can be qualitatively explained as follows. 
First, remind that the modes $\hat{c}_{1}$ and $\hat{d}_{1}$ share correlated noises. 
This quantum correlation is then transferred so that the modes $\hat{a}_{2}$ and $\hat{c}_{1}$ become correlated after the combination of $\hat{a}_{1}$ and $\hat{d}_{1}$ performed in BS. 
In final, interference-induced quantum noise cancellation occurs as the internal degree of amplifier (i.e., FWM$_{2}$) is correlated with the input signal mode~\cite{Kong2013}. 
Therefore, an ACNC is established. 
We can use the ACNC to recover the quantum information of the state of the initial input mode \(a_1\).

\section{Recovering quantum coherence via ACNC} \label{sec:recovery_coherence}

\subsection{Coherence measure for Gaussian states}

We use quantum coherence~\cite{PhysRevLett.113.140401,RevModPhys.89.041003} as an evaluating indicator to quantitatively investigate the recovery capability of the ACNC for the continuous-variable quantum information. 
We shall give a brief review on the quantum coherence of Gaussian states.
Baumgratz, Cramer, and Plenio defined the quantum coherence of a quantum state \(\hat\rho\) as the minimum distance measured by the quantum relative entropy between the quantum state and an incoherent state in the Hilbert space~\cite{PhysRevLett.113.140401}.
Denote by \(\mathcal I\) the set of all incoherent states whose density matrices are diagonal in the fixed reference basis.
The relative entropy of coherence is defined as~\cite{PhysRevLett.113.140401}
\begin{equation}
	C_\mathrm{r}(\hat\rho) = \min_{\hat\sigma\in \mathcal I} S(\hat\rho || \hat\sigma),
\end{equation}
where \(S(\hat\rho || \hat\sigma) = \tr(\hat\rho \log_2 \hat\rho) - \tr(\hat\rho\log_2\hat\sigma)\) is the quantum relative entropy between \(\hat\rho\) and \(\hat\sigma\).
The relative entropy of coherence can be expressed as
\begin{equation}
	C_\mathrm{r}(\hat\rho) 
	= S(\hat\rho_\mathrm{diag}) - S(\hat\rho),
\end{equation}
where $S(\hat\rho) = - \tr(\hat\rho\log_2\hat\rho)$ is the von Neumann entropy of \(\hat\rho\) and $\hat{\rho}_{\rm{diag}}$ denote the diagonal matrix obtained by removing all off-diagonal elements from $\hat{\rho}$ in the reference basis~\cite{PhysRevLett.113.140401}.
The relative entropy of coherence also serves as a well-defined quantifiers for quantum coherence in infinite-dimensional bosonic systems~\cite{PhysRevA.93.012334}. 

For Gaussian states of a bosonic system, Xu~\cite{PhysRevA.93.032111} gave an alternative coherence measure---the Gaussian relative entropy of coherence:
\begin{equation}
	C(\hat\rho) = \min_{\hat\sigma\in \mathcal I'} S(\hat\rho || \hat\sigma),
\end{equation}
where \(\mathcal I'\) denotes the set of all incoherent Gaussian states with respect to the multimode Fock basis.
Moreover, Xu~\cite{PhysRevA.93.032111} showed that the closest incoherent Gaussian state to a \(N\)-mode Gaussian state \(\hat\rho\) is the \(N\)-mode thermal state
\begin{equation}
	\hat{\rho}_\mathrm{th} 
	= \bigotimes^N_{j=1} \qty[\sum_{n=0}^{\infty}\frac{\bar{n}^n_j}{(\Bar{n}_j+1)^{n+1}}\op{n}],
\end{equation}
where $\bar{n}_j$ is the mean number of photons in the $\textit{j}$-th mode. 
Therefore, the Gaussian relative entropy of coherence can be expressed as
\begin{equation}\label{relativeentropy}
    C(\hat\rho) 
    = S(\hat\rho_\mathrm{th}) - S(\hat\rho).
\end{equation}
The von Neumann entropy of \(N\)-mode thermal state can be directly calculated as
\begin{equation} \label{eq:von_entropy_thermal}
	S(\hat\rho_\mathrm{th}) 
	= \sum_{j=1}^N \qty[
		(\bar{n}_j + 1) \log_2(\bar{n}_j + 1) - \bar{n}_j \log_{2} \bar{n}_j
	].
\end{equation}
The von Neumann entropy of \(N\)-mode Gaussian states can be expressed as
\begin{equation} \label{eq:gaussian_entropy}
	S(\hat\rho)
	= \sum^N \limits_{j=1} \qty(
		\frac{\nu_j + 1}{2}\log_2\frac{\nu_{\emph{j}}+1}{2}
		- \frac{\nu_{\emph{j}}-1}{2} \log_2\frac{\nu_{\emph{j}}-1}{2}
	),
\end{equation}
where $\nu_{j}$ is the symplectic eigenvalue of the covariance matrix $V$~\cite{Holevo1999,RevModPhys.84.621}.
The entries of \(V\) are defined by $V_{ij} = \frac{1}{2}\langle\hat{x}_{i}\hat{x}_{j}+\hat{x}_{j}\hat{x}_{i}\rangle-\langle\hat{x}_{i}\rangle\langle\hat{x}_{j}\rangle$, where $\langle\bullet\rangle$ denotes the quantum expectation and $\hat{x} = (\hat{X}_{1},\hat{P}_{1}, \ldots,\hat{X}_{N},\hat{P}_{N})^{T}$ with $\hat{X}_{j}=\hat{a}_{j}+\hat{a}^{\dagger}_{j}$ and $\hat{P}_{j}=i(\hat{a}^{\dagger}_{j}-\hat{a}_{j})$. 
It can be computed as the absolute value of the eigenvalues of the matrix $i\Omega V$, where $\Omega=\bigoplus^{N}_{k=1} \smqty( 0 &  1 \\ -1 &  0)$ is the symplectic transformation matrix.

\subsection{Recovering quantum coherence of a coherent state}

\begin{figure}[tb]
\centering
\includegraphics[width=0.8\linewidth]{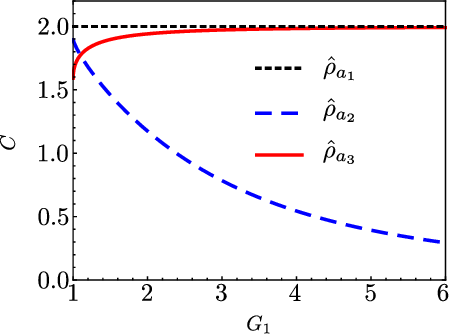}
\caption{
	Quantum coherence of a coherent state transmitted in quantum channels. 
	Here, $\hat{\rho}_{a_{1}}$ is the initial state, \(\rho_{a_2}\) is the output state of the first noisy channel modeled by a beam splitter, and \(\rho_{a_3}\) is the output state of the ACNC. 
	The parameters in plotting this figure are $\alpha=1$, $G_{2}=1/\sqrt{T}$, and $T=0.9$.}
\label{fig2}
\end{figure}

We now consider the capability of the ACNC in recovering the quantum coherence when the input state of \(a_1\) mode is a coherent state \(\ket\alpha\) and the states of \(c_0\) and \(d_0\) are the vacuum states.
Let us denote by \(\hat\rho_{a_1}\), \(\hat\rho_{a_2}\), and \(\hat\rho_{a_3}\) the quantum states of the modes \(\hat a_1\), \(\hat a_2\), and \(\hat a_3\), respectively.
These states are all Gaussian, so we can use Eqs.~(\ref{relativeentropy}--\ref{eq:gaussian_entropy}) with Eqs.~(\ref{a_2}) and (\ref{acncoutput2}) to calculate their Gaussian relative entropy of coherence.
Since the input state \(\ket\alpha\) is a pure state whose von Neumann entropy vanishes, the quantum coherence of the input state is just \(S(\hat\rho_\mathrm{th})\) given by Eq.~\eqref{eq:von_entropy_thermal} with \(N=1\) (viz., the single mode case) and \(n_1=|\alpha|^2\).
After the noisy channel modeled by the BS, the mean photon number is 
\begin{equation}
	\ev*{\hat a_2^\dagger \hat a_2} = T|\alpha|^2 + (1-T) g_1^2,
\end{equation}
from which we can obtain \(S(\hat\rho_\mathrm{th})\) via Eq.~\eqref{eq:von_entropy_thermal}.
Meanwhile, the covariance matrix for \((\hat X_2, \hat P_2)\) is given by 
\begin{align}
	V_{11} &= V_{22} = 1+2(1-T)g_1^2, \\
	V_{12} &= V_{21} = 0.
\end{align}
The symplectic eigenvalue of the covariance matrix is \(\nu_1 = 1+2(1-T)g_1^2\), with which we can obtain the von Neumann entropy of \(\hat\rho_{a_2}\).
After the ACNC, the mean photon number in the \(\hat a_3\) mode is 
\begin{equation}
	\ev*{\hat a_3^\dagger \hat a_3}
	= |\alpha|^2+g_2^2 (G_1 - g_1)^2
\end{equation}
and the covariance matrix for \((\hat X_3, \hat P_3)\) is given by 
\begin{align}
	V_{11} &= V_{22} = 1+2g_{2}^{2}\left(G_{1}-g_{1}\right)^{2}, \\
	V_{12} &= V_{21} = 0.
\end{align}

As shown in Fig.~\ref{fig2}, the quantum coherence of \(\hat \rho_{a_2}\) is smaller than that of \(\hat\rho_{a_1}\) and decreases with the increase of $G_{1}$. 
It means that the noisy channel destroys the quantum coherence of $\hat{a}_{1}$ and the decoherence becomes more obvious with the increase of the thermal noise introduced. 
On the other hand, the quantum coherence of \(\hat\rho_{a_2}\), which is the output state of the ACNC, can be partially recovered and approaches to the quantum coherence of \(\hat\rho_{a_1}\) with increasing $G_{1}$. 
In other words, the decoherence caused by the noisy channel can be mitigated by the ACNC. When $G_{1}\rightarrow\infty$, the quantum coherence of $\hat{a}_{1}$ can be totally recovered.

\begin{figure}[tb]
\centering
\includegraphics[width=0.8\linewidth]{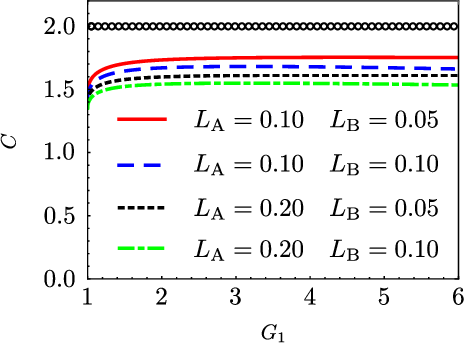}
\caption{\label{fig3}
	Quantum coherence of the mode \(\hat a_3\) versus \(G_1\) under various losses.
	The circles denote the quantum coherence of the initial coherent state $\hat{\rho}_{a_{1}}$ for the convenience of comparison. 
	The parameters in plotting this figure are $\alpha=1$, $G_{2}=1/\sqrt{T}$, and $T=0.9$.
}
\end{figure}

\emph{Effect of losses.} 
In the above discussions, we have introduced two types of quantum channels, that is, the noisy channel and the ACNC. 
The noisy channel results in decoherence of the input Gaussian state, while the ACNC is capable of recovering the quantum coherence of the state that is degraded by the noisy channel. 
In the realistic scenario, the performance of the ACNC will be affected by unavoidable losses, such as atomic absorption during the FWM processes and losses in light paths~\cite{PhysRevA.78.043816,XinJun2017a}. 
In the following, we study how the losses affect the performance of the ACNC on coherence recovery. 
We model the lossy process as a BS with transmissivity $\eta$ and use $L \equiv 1-\eta$ to quantify the strength of loss. 
When an optical field $\hat{a}_{j}$ passes through the lossy channel, it becomes $\hat{a}_{j}\rightarrow\sqrt{\eta_{j}}\hat{a}_{j}+\sqrt{1-\eta_{j}}\hat{v}_{j}$, where $\hat{v}_{j}$ is the annihilation operator for the other input port of the BS.
The state of the mode \(\hat v_j\) is the vacuum state.

To describe the atomic absorption during the FWM processes, we assume that each of the modes $\hat{d}_{1}$, $\hat{c}_{1}$, and $\hat{a}_{3}$, which are the relevant outputs of the involved FWMs, undergoes a lossy channel with the strength $L_{\rm{A}}$. 
We also assume that there is a photon loss with the strength $L_{\rm{B}}$ occurring in the propagating mode $\hat{a}_{2}$. 
We consider the situations where the losses caused by atomic absorption is of $10\%$--$20\%$ and the losses in light path is of $5\%$--$10\%$, as they give realistic experimental losses~\cite{XinJun2017a}. 
We plot in Fig.~\ref{fig3} the quantum coherence of the mode $\hat{a}_{3}$ versus $G_{1}$ for different loss strengths.
Our results show that the ACNC cannot totally recover the quantum coherence of the mode $\hat{a}_{3}$ to that of the input mode $\hat{a}_{1}$ (which is of value $2$) in the presence of losses. 
This is because the extra noise added by photon loss is uncorrelated. 
Different from the correlated noisy channel, the quantum decoherence caused by photon loss cannot be mitigated in principle.

\begin{figure}[tb]
	\centering
	\includegraphics[width=3.4in]{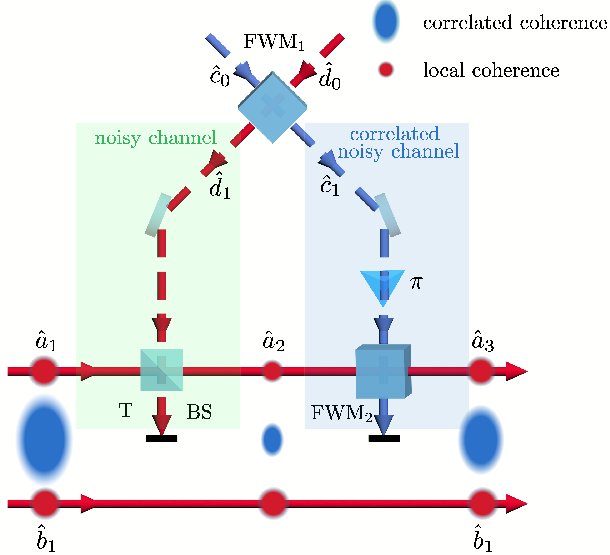}
	\caption{Recovering quantum coherence of a TMSS via ACNC.}
	\label{fig4}
\end{figure}

\begin{figure*}[tb]
	\centering
	\includegraphics[width=6.8in]{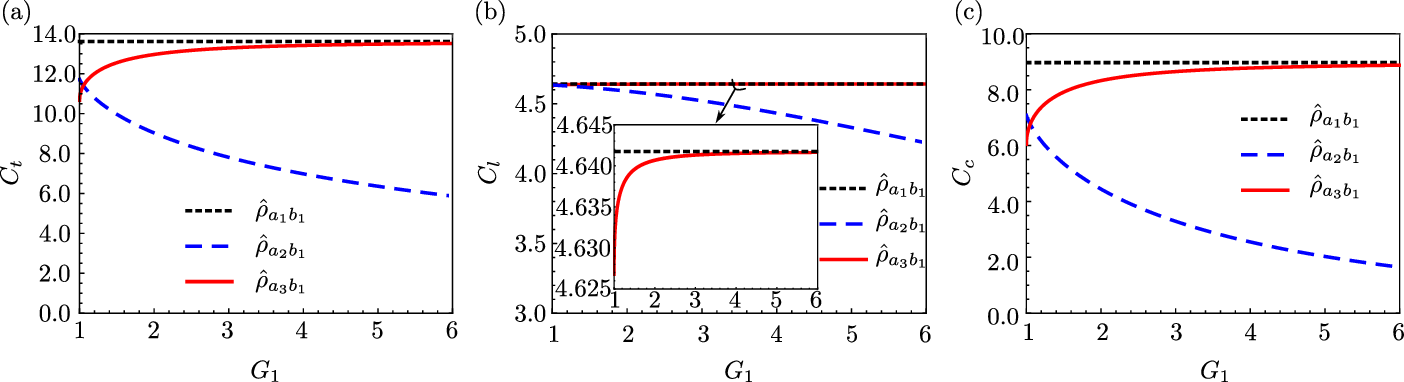}
	\caption{Different kinds of quantum coherence of a TMSS transmitted via the noise channels and its ACNC. 
	(a) Total quantum coherence; (b) Local quantum coherence; (c) Correlated quantum coherence. 
	The black dotted line corresponds to the quantum coherence of the initial TMSS $\hat{\rho}_{a_{1}b_{1}}$. 
	The parameters in plotting this figure are \(\theta = 0\), $G_0 = 3$, $G_2 = 1 / \sqrt{T}$, and $T=0.9$.
	The initial state of the FWM that generates the TMSS is \(\ket\alpha \otimes \ket\alpha\) with \(\alpha = 1 \).
	}
	\label{fig5}
\end{figure*}

\subsection{Recovering quantum coherence of a TMSS}
In the above, we have shown that the ACNC is able to recover the quantum coherence of coherent states undergoing a noisy channel. 
We now generalize our model to bipartite quantum systems. 
For a two-mode quantum state $\hat{\rho}_{\rm{AB}}$, its total quantum coherence \(C_t(\hat\rho_\mathrm{AB}) \equiv \mathcal C(\hat\rho_\mathrm{AB})\) can be decomposed into two parts: the local coherence and the correlated coherence~\cite{PhysRevA.94.022329}. 
The local coherence of $\hat{\rho}_{\rm{AB}}$ is defined as
\begin{equation}
	C_{l}(\hat{\rho}_{\rm{AB}})
	\equiv C(\hat{\rho}_{\rm{A}}) + C(\hat{\rho}_{\rm{B}}),
\end{equation}
where $C(\hat{\rho}_{\rm{A}})$ and $C(\hat{\rho}_{\rm{B}})$ are quantum coherence of the reduced density operators of the modes A and B, respectively. 
In general, it is not necessary that all quantum coherence of a bipartite quantum system are stored locally. 
A part of quantum coherence may be stored in the correlation between the subsystems. 
The difference between total coherence $C_{t}(\hat{\rho}_{\rm{AB}})$ and local coherence $C_{l}(\hat{\rho}_{\rm{AB}})$ is defined as correlated coherence~\cite{PhysRevA.94.022329}, which is denoted by $C_{c}(\hat{\rho}_{\rm{AB}})$, i.e.,
\begin{equation}
	C_{c}(\hat{\rho}_{\rm{AB}})
	\equiv C_{t}(\hat{\rho}_{\rm{AB}})-C_{l}(\hat{\rho}_{\rm{AB}}).
\end{equation}

As shown in Fig.~\ref{fig4}, we consider a TMSS denoted by $\hat{\rho}_{a_{1}b_{1}}$ as the input state of the noisy channel. 
In experiment, the quantum state $\hat{\rho}_{a_{1}b_{1}}$ can be generated by a two-beam phase sensitive FWM process~\cite{PhysRevLett.123.113602} with the input-output relation
\begin{eqnarray}
	\nonumber
	\hat{a}_{1}&=&G_{0}\hat{a}_{0}+e^{i\theta}g_{0}\hat{b}^{\dagger}_{0}, \\
	\hat{b}_{1}&=&G_{0}\hat{b}_{0}+e^{i\theta}g_{0}\hat{a}^{\dagger}_{0},
\end{eqnarray}
where $\hat{a}_{0}$ and $\hat{b}_{0}$ are the input modes of the FWM process and are both assumed to be in coherent states, the parameter $\theta$ denotes the phase of the two-beam phase sensitive FWM process, $G_{0}$ is the amplitude gain, and $g_{0} \equiv \sqrt{G^{2}_{0}-1}$. 
It has been shown that interference-induced quantum squeezing can be achieved by such a TMSS.
Moreover, the quantum squeezing reaches its maximum when $\theta=0$, corresponding to the bright interference fringe of the output ports of the FWM process.

To study the performance of the ACNC in recovering the quantum coherence of a TMSS, we first seed mode $\hat{a}_{1}$ into the noisy channel whose output mode is denoted by $\hat{a}_{2}$. 
After the noisy channel, we seed mode $\hat{a}_{2}$ into the ACNC. 
Similar with the single-mode case as shown in Fig.~\ref{fig1}, the noisy channel is realized by mixing mode $\hat{a}_{1}$ with $\hat{d}_{1}$ and the associated ACNC is realized by FWM$_{2}$, which uses correlated modes $\hat{a}_{2}$ and $\hat{c}_{1}$ as its input. 
The output of $\hat{a}_3$ can be written as 
\begin{equation}
	\hat{a}_{3} = G_{0}\hat{a}_{0}+e^{i\theta}g_{0}\hat{b}^{\dagger}_{0}+g_{2}(G_{1}-g_{1})(\hat{d}_{0}-\hat{c}^{\dagger}_{0}).
\end{equation}

We plot in Fig.~\ref{fig5} the total quantum coherence, the local quantum coherence, and the correlated quantum coherence for the bipartite state at different stages.
% (a) , where black dotted, blue dashed and red solid lines corresponds to the coherence of quantum state $\hat{\rho}_{a_{1}b_{1}}$, $\hat{\rho}_{a_{2}b_{1}}$ and $\hat{\rho}_{a_{3}b_{1}}$.
% We first focus on the total quantum coherence 
As shown in Fig.~\ref{fig5} (a), the effects of the noisy channel and the ACNC on the total quantum coherence of the TMSS are similar with that of the single-mode coherent state, which is shown in Fig.~\ref{fig2}.
It demonstrates that the ACNC can recover not only the quantum coherence of a single-mode coherent state undergoing the noisy channel but also the total quantum coherence of a bipartite system whose subsystem undergoing the noisy channel.
% The quantum coherence of the TMSS can also be recovered by the ACNC.
Comparing Fig.~\ref{fig5}~(b) and Fig.~\ref{fig5}~(c), it can be seen that the local coherence of the TMSS is more robust against the noisy channel than the correlated coherence.
Moreover, the local coherence of the TMSS can be almost recovered by the ACNC as long as $G_{1}>1$; the correlated coherence of $\hat{\rho}_{a_{3}b_{1}}$, however, can approach to that of the input state $\hat{\rho}_{a_{1}b_{1}}$ only if $G_{1}$ is large enough.

\begin{figure*}[tb]
	\centering
	\includegraphics[width=6.8in]{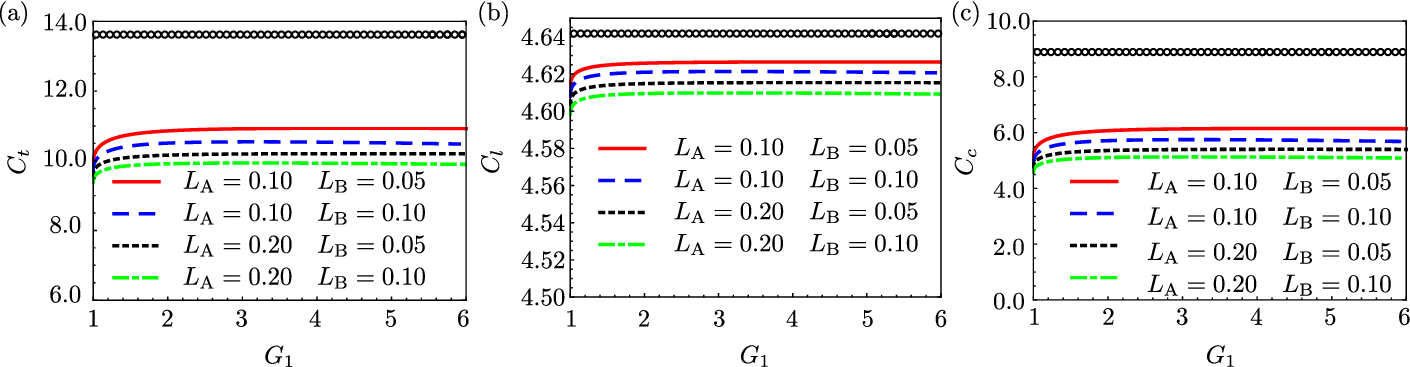}
	\caption{Effect of losses on the performance of the ACNC in recovering the quantum coherence of a TMSS. 
	(a) Total quantum coherence; (b) Local quantum coherence; (c) Correlated quantum coherence. 
	The circle indicates the quantum coherence of the initial TMSS $\hat{\rho}_{a_{1}b_{1}}$ for the convenience of comparison. 
	The parameters in plotting this figure are \(\theta = 0\), $G_0 = 3$, $G_2 = 1 / \sqrt{T}$, and $T=0.9$.
	The initial state of the FWM that generates the TMSS is \(\ket\alpha \otimes \ket\alpha\) with \(\alpha = 1\).
	}
	\label{fig6}
\end{figure*}

\emph{Effect of losses.} 
We show in Fig.~\ref{fig6} the effect of losses on the performance of the ACNC in recovering the quantum coherence of the TMSS. 
We first focus on the local coherence. 
As shown in Fig.~\ref{fig6}~(b), the local coherence for different lossy cases are almost the same and all of them are very close to that of the input state $\hat{\rho}_{a_{1}b_{1}}$. 
It means that the local coherence of the TMSS owns good robustness against losses. 
Different from the local coherence, the correlated coherence is vulnerable to the losses. 
As shown in Fig.~\ref{fig6}~(c), the correlated coherence of the state $\hat{\rho}_{a_{3}b_{1}}$ decreases rapidly with the increase of the losses.
The ACNC can never recover the quantum coherence of the state $\hat{\rho}_{a_{3}b_{1}}$ to that of the input state $\hat{\rho}_{a_{1}b_{1}}$ as long as losses are involved.

\section{Conclusion} \label{sec:conclusion}

In this work, we have proposed an ACNC relying on FWM processes and showed that the ACNC can utilize the correlated noise to recover a portion of coherent loss while amplifying the complex amplitude to compensate the attenuation.
The protocol has good performance for not only single-mode coherent states but also the TMSS. 
By dividing the total coherence of the TMSS into the local coherence and the correlated coherence, we find that the local coherence is more robustness against thermal noise than the correlated coherence. 
We have also investigated the effect of imperfect factors on the performance of the ACNC, such as the photon absorption occurring in the FWM processes and unavoidable losses in light path. 

The ACNC has some advantages for recovering quantum coherence.
First, its capability for coherence recovery is universal for any input quantum state.
This is because the recovery capability is based on the operator level in the Heisenberg picture so the mechanism of recovering coherence is irrelevant to the input states.
Second, the ACNC could own larger operational bandwidth, compared with the conventional correlated noisy channels proposed and experimentally in Refs.~\cite{PhysRevLett.111.180502,Deng2016a,Deng2021}, which are limited by the electrical bandwidth of the electro-optic modulators used therein. 
Different from these previous works, the ACNC is all optical so that it avoids the electro-optic conversion. 
The ACNC proposed in this work is hopeful of being used to study the decoherence effect in quantum optics.

\begin{acknowledgments}
This work is supported by the National Natural Science Foundation of China (Grants No.\ 12275062, No.\ 11935012, and No.\ 61871162).
\end{acknowledgments}

\bibliography{reference}

\end{document}